\documentclass[11pt]{article}
\usepackage{jheppub}
\pdfoutput=1

\usepackage{slashed}
\usepackage{array,multirow}
\usepackage{graphicx,graphics}
\usepackage{amsfonts,amsmath,amssymb, bbold}
\usepackage{color,xcolor}
\usepackage{tikz} 
\usepackage{enumitem}
\usepackage{cleveref}
\usepackage{subcaption}

\newcommand{\be}{\begin{equation}}
\newcommand{\ee}{\end{equation}}
\newcommand{\eps}{\epsilon}

\renewcommand\({\left(}
\renewcommand\){\right)}
\renewcommand\[{\left[}
\renewcommand\]{\right]}

\newcommand{\dd}{{\rm d}}
\newcommand{\e}{{\rm e}}
\newcommand{\tr}{{\rm tr}}

\def\nn{\nonumber}

\title{Cosmological phase transitions: is effective field theory just a toy?
}

\vspace{2.4cm}

\abstract{ To obtain a first order phase transition requires large new
  physics corrections to the Standard Model (SM) Higgs potential. This
  implies that the scale of new physics is relatively low, raising the
  question whether an effective field theory (EFT) description can be
  used to analyse the phase transition in a (nearly) model-independent
  way.  We show analytically and numerically that first order phase
  transitions in perturbative extensions of the SM cannot be described
  by the SM-EFT. The exception are Higgs-singlet extension with tree-level
  matching; but even in this case the SM-EFT can only capture part of
  the full parameter space, and if truncated at dim-6 operators, the
  description is at most qualitative. We also comment on the
  applicability of EFT techniques to dark sector phase transitions.  }

\author[a]{Marieke Postma} \author[b]{and Graham White}
\emailAdd{mpostma@nikhef.nl}
\emailAdd{graham.white@ipmu.jp}
\affiliation[a]{Nikhef,
Science Park 105,
1098 XG Amsterdam, The Netherlands}
\affiliation[b]{Kavli IPMU (WPI), UTIAS, The University of Tokyo, Kashiwa, Chiba 277-8583, Japan}

\preprint{Nikhef 2020-038}

\begin{document}

\maketitle

\section{Introduction}

Determining the nature of the electroweak phase transition would be a
major scientific achievement. The Standard Model (SM) predicts a
crossover, but the transition may be different in extensions of the SM provided the new physics is important at the electroweak scale.  Such new physics can be searched for at the LHC and, with more precision, in next generation colliders \cite{Ramsey-Musolf:2019lsf,Kotwal:2016tex,Huang:2017jws,
  Benedikt:2653674,Papaefstathiou:2020iag}. Especially interesting is the possibility of a strongly first-order electroweak phase transition (SFO-EWPT). The ensuing bubble dynamics could provide suitable conditions for producing the observed asymmetry between baryons and anti-baryons \cite{Kuzmin:1985mm,Shaposhnikov:1986jp,Shaposhnikov:1987tw,Cohen:1993nk} (see \cite{Morrissey:2012db,White:2016nbo} for a review), and moreover can produce
a potentially observable stochastic background of gravitational waves in the frequency range that LISA will be sensitive to \cite{Caprini:2019egz}.

The above considerations have motivated the construction of many SM
extensions with a SFO-EWPT.  It would be advantageous if these could
be studied in a single framework, allowing for a (nearly)
model-independent assessment of key aspects, such as the strength of
the phase transition and the phenomenogical implications.  The
Standard Model effective field theory (SMEFT)
\cite{Jenkins:2013zja,Brivio:2017vri} may provide such an
approach. The idea is that the new physics degrees of freedom are
heavy and can be integrated out; their effects on the low-energy SM
degrees of freedom are then parameterized by a tower of higher
dimensional operators. If there is a sufficient separation of scales
between the light and heavy fields, then the higher the mass dimension
of the operator the more suppressed the impact on low energy
observables, and consequently the EFT can be truncated at a given
dimension depending on the desired precision.  The EFT approach to the
electroweak phase transition has been explored in multiple studies
\cite{Grojean:2004xa,Bodeker:2004ws,Delaunay:2007wb,Balazs:2016yvi,deVries:2017ncy,deVries:2018tgs,Chala:2018ari,Ellis:2019flb,Croon:2020cgk}.

Lattice calculations show that in the standard model, a SFO-EWPT is
only achieved for a Higgs mass $m_{h0} \lesssim 65\,$GeV
\cite{Kajantie:1995kf,Kajantie:1996mn,Kajantie:1996qd,Csikor:1998eu,DOnofrio:2015gop}
well below the measured value $m_{h0} \approxeq 125.1\,$GeV
\cite{Tanabashi:2018oca}. To obtain a strong phase transition
therefore requires a large modification of the SM Higgs potential, by
order one effects, which can only be achieved with new degrees of
freedom that are sufficiently light.  A first order phase transition
requires a barrier (at finite temperature) between the false vacuum at
the origin and the electroweak vacuum at non-zero Higgs field
values. As was noted in \cite{deVries:2017ncy}, in the SMEFT, the
local maximum follows from balancing the quadratic and (negative)
quartic terms in the potential, whereas the minimum at finite vacuum
expectation value (vev) is obtained balancing the quartic term with
the higher dimensional operators.  This implies that there is no separation
of scales, and one generically expects the EFT approach to break down
-- this was indeed observed in the specific set-up of
\cite{deVries:2017ncy}, where it was found that the effect of
dimension 8 operators could be as large as the dimension 6
operators. Ref. \cite{Damgaard:2015con} did a numerical analysis
comparing the Higgs-singlet model with the SMEFT approximation, and
also found the EFT does not provide a good description.

In this work we perform a systematic study of the validity of the
SMEFT description to capture perturbative UV models with a strongly
first-order electroweak phase transition. In matching the UV theory to
SMEFT, the Wilson coefficients of the non-renormalizable operators
can be generated at tree and/or loop level. We find:
\begin{itemize}
\item In set-ups with only loop level matching, the SMEFT expansion
  breaks down, and the EFT cannot be truncated at operators of a given
  mass dimension.
\item In set-ups with tree level matching, the SMEFT expansion also
  breaks down, with the possible exception of Higgs-singlet extensions.
\item In Higgs-singlet extension with tree level matching,
  i.e. without a $Z_2$ symmetry, the SMEFT description is (marginally)
  valid only in part of the parameter space for a SFO-EWPT. For accurate
  results dimension 8 operators need to be included, even though the impact
  of dimension 10 and higher order operators may be small.
\end{itemize}  
As the SMEFT can only be used for a single SM extension, and with
limited success, there is evidently no advantage in using the EFT
approach over studying the UV set-up itself.  There are many papers
studying the electroweak phase transition and the implications for
baryogenesis and gravitational wave production using SMEFT with
dimension 6 operators\cite{Grojean:2004xa,Delaunay:2007wb,Damgaard:2015con,Balazs:2016yvi,deVries:2017ncy,deVries:2018tgs,Chala:2018ari,Ellis:2019flb,Phong:2020ybr}, partially because it is very
tractable. Our results imply that for most of the interesting
parameter space no UV completion exists, and for those points that can
be mapped to a Higgs-singlet model, only qualitative results can be
obtained.

Finally we note that the EFT language is also used to describe
strongly first order phase transitions (SFO-PTs) in a hidden sector, to
determine the produced background of gravitational waves
\cite{Baldes:2017rcu,Croon:2018erz,Croon:2019rqu}. As the dark sector
is relatively unconstrained, e.g. the mass of the dark Higgs and its
couplings to dark fermions and gauge bosons are unknown, it is not
surprising that the separation of scales required for the validity of
the EFT expansion can be achieved.  Nonetheless, we can also
formulate conditions on the validity (and usefulness) of the EFT
framework for dark sector SFO-PTs.

The structure of this paper is as follows. In \cref{sec:SMEFT} we
introduce the Standard Model effective field theory and we discuss the
requirements for a first order phase transion, the thermal corrections
to the Lagrangian in the early universe, the validity of the SMEFT
expansion, and the generalization to dark sectors. In
\cref{sec:matching} we review the matching results at tree and loop
level. We focus on SM extensions with additional scalars (or gauge
bosons), as these can facilitate a first order phase transition.  We
then discuss the implications for the EWPT and the validity of the SMEFT
expansion in \cref{sec:explicitmatch}.  As the singlet-Higgs extension
is the most interesting in this context, we provide numerical results
for this set-up as well. Details on the numerical implementation can
be found in \cref{A:scan}. We end this section with some comments on
dark sector phase transitions. Our results are summarized in 
\Cref{sec:conclusion}.

\section{SMEFT and first order phase transitions}\label{sec:SMEFT}

The SM effective field theory (SMEFT) rests on the assumption that the
new particles in extensions of the SM have a mass larger than the
electroweak scale, i.e. the scale of the SM states.  The effective
theory at the electroweak scale is then the SM augmented with a series
of gauge invariant higher dimensional operators constructed out of the
SM fields, to incorporate the effects of integrating out the heavy physics:
\be
{\cal L}_{\rm SMEFT} = {\cal L}_{\rm SM} + \sum_i \frac1{M^{d_i-4}}c_i {\cal O}_i, \qquad
{\cal L}_{\rm SM}=|D_\mu H|^2- (\mu^2 |H|^2 +\lambda |H|^4)+ ...
\ee
Here $M$ is the mass scale of the heavy particles, $c_i$ the Wilson coefficients and $d_i$ the mass dimension of the operator ${\cal O}_i $. For the electroweak phase transition we are interested in corrections to the kinetic term and potential for the SM Higgs doublet $H$, and we only consider operators $ {\cal O}_i= {\cal O}_i(H)$.

In this section we discuss the Higgs potential in SMEFT with only
dimension 6 operators included, and identify the conditions for a
strongly first order electroweak phase transition and the validity of
the EFT expansion.

%%%%%%%%%%%%%%%%%%%%%%%%%%%%%%%%%%%%%%%%%%%%%%%%%
\subsection{The Higgs potential in SMEFT}

The operators in the SMEFT that are at most dimension 6 and relevant for Higgs dynamics are (listed in the Warsaw basis \cite{Grzadkowski:2010es})
\begin{align}
  {\cal L}^{(6)}_{\rm SMEFT} &= c_{\rm H\Box} |H|^2 \Box |H|^2 +
  c_{HD}|HD_\mu H|^2 + c_H |H|^6+ {\cal   O}(M^{-4})
\nn \\ &
= c_{\rm kin} \bar h^2 (\partial \bar h)^2 + \frac18 c_H \bar h^6
\label{L_wilson}
\end{align}  
with the last expression written in unitary gauge $\sqrt{2}H^\top= (0
\;\bar h)$, and $ c_{\rm kin} =\frac14 c_{HD} -c_{\rm H\Box} $.  We
can define the approximate canonical field $h =\bar h +\frac13 c_{\rm
  kin}\bar h^3 +{\cal O} (c_{\rm kin}^2 \bar h^5)$, solve for $\bar
h(h)$ and write the Higgs Lagrangian as
\be
{\cal L}_{\rm SMEFT}^{(6)} \simeq \frac12 (\partial h)^2- \(
\frac12 a_2 h^2 +\frac14 a_4 h^4 + \frac16 a_6 h^6 \),
\label{Va}
\ee
with
\begin{align}
a_2 =\mu^2, \quad
      a_4 =\lambda- \frac43  c_{\rm kin} \mu^2,\quad
            %=\lambda-\frac13 \mu^2 (c_{HD} -4c_{\rm H\Box}),\nn \\
  a_6= -\frac{3}{4}c_H-2  c_{\rm kin} \lambda.
  % =- \frac14\(3c_H+2  \lambda(c_{HD} -4c_{\rm H\Box})\)
  \label{akin}
\end{align}
The parameters $a_2,a_4$ are fixed by the measured Higgs vev $v=246\,$GeV and Higgs mass $m_{h0}=125\,$GeV via
\begin{equation}
  \label{higgs_mass}
  \partial_hV|_{h=v} =0, \quad\partial_h^2 V|_{h=v} =m_{h0}^2.
\end{equation}
Rewriting the tree-level potential in terms of these physical quantities gives
\be
V = -\frac14(m_{h0}^2 -2 a_6 v^4) h^2 +\frac14\(\frac{m_{h0}^2}{2v^2} -2 a_6 v^2\)h^4 +\frac16 a_6 h^6.
\label{V_phys}
\ee
%

% The full potential governing the phase transition includes the one-loop Coleman-Weinberg (CW) contribution \cite{Coleman:1973jx} and the thermal corrections of the SM particles $V_{\rm eff} = V + V_{\rm CW}+ V_T$. The effective action in the loop expansion is not gauge invariant \cite{Jackiw:1974cv,Patel:2011th}\footnote{Although the gauge dependence cancels at one-loop order when working with the canonical Higgs field \cite{Frere:1974ia,Sher:1983em,Espinosa:2015qea} as a consequence of the Nielsen identities.\cite{Nielsen:1975fs}} and the CW potential introduces a scale dependence; if the theoretical uncertainty this causes \cite{Croon:2020cgk} becomes large, higher order loop contributions should be included. Given the fact that the scale dependence is the dominant error and the $\hbar$ expansion only allows a resummation at ${\cal O} (\hbar ^2)$, we choose to work with a loop expansion including Arnold Espinosa resummation \cite{Arnold:1992rz} in the On shell renormalization scheme \cite{Delaunay:2007wb,Curtin:2014jma} with counter terms chosen such that $\partial_h V_{\rm CW}|_{h=v}=\partial^2_h V_{\rm CW}|_{h=v}=0$ at the $Z$-pole scale.

The full potential governing the phase transition includes the
one-loop Coleman-Weinberg (CW) contribution \cite{Coleman:1973jx} and
the thermal corrections of the SM particles
$V_{\rm eff} = V + V_{\rm CW}+ V_T$. The  off-shell effective action
is gauge dependent \cite{Jackiw:1974cv,Patel:2011th}, but at one loop
order the gauge dependence is cancelled when rewritten in terms of the
canonical Higgs field 
\cite{Frere:1974ia,Sher:1983em,Espinosa:2015qea}; another prescription
to deal with the gauge dependence can be found in
\cite{Patel:2011th}. The CW potential introduces a scale dependence;
if the theoretical uncertainty this causes 
becomes large \cite{Croon:2020cgk}, higher order loop contributions should be
included. That being said, we choose to work in the on shell
renormalization scheme \cite{Delaunay:2007wb,Curtin:2014jma} with
counter terms chosen such that
$\partial_h V_{\rm CW}|_{h=v}=\partial^2_h V_{\rm CW}|_{h=v}=0$ at the
$Z$-pole scale. This results in the property that the higgs vev and
mass are set by the parameters of the tree-level potential, which is
very convenient for numerical scans. In this prescription, the
one-loop Coleman-Weinberg potential is given by
\be V_{\rm CW} =
\sum_i \frac{n_i}{(8\pi)^2}
\[ m_i^4\(\ln \(\frac{m_i^2}{m_{0i}^2}\)-\frac32\)+2m_i^2m_{0i}^2\],
\ee
with $m_{i}$ the field-dependent mass, $m_{0i}$ the vacuum mass at
$h=v$, and $n_i =\{1,3,3,6,-12\}$ the degrees of freedom (d.o.f). of
the Higgs, goldstones, $Z$, $W$ and top quark respectively, which give
the dominant contributions.  The thermal potential is
\begin{align}
  V_T &= \sum_{i={\rm bosons }} n_i T^4 J_B (\frac{m_{h}^2}{T^2})  +\sum_{j={\rm fermions }} n_j  T^4 J_F (\frac{m_{h}^2}{T^2}) 
        \label{VT}
\end{align}
with $n_i,n_j$ the bosonic and fermion degrees of freedom, and the
explicit thermal functions $J_{B/F}$  given in \cref{J_lim}. For
the bosonic and longitudinal gauge d.o.f. we include the infrared
contribution from daisy diagrams
\cite{Dolan:1973qd,Carrington:1991hz}. To leading order in the
high-temperature expansion, giving $V_{\rm eff}$ up to ${\cal O}(T^0)$
corrections, this is equivalent to replacing
$m^2_i \to m_i^2 + \Pi_{i}$ with $\Pi_i$ thermal self energies
\cite{Arnold:1992rz}. More details can be found in \cref{A:scan}.
We expect that our main (qualitative) results on the validity of the EFT description
for a SFO-EWPT will not depend on the details of how renormalization and
thermal resummations are implemented.

% Finally, we resum the bosonic diagrams through the usual Arnold-Espinosa resummation scheme which results in the term \cite{Arnold:1992rz}
% \begin{equation}
%  V_D = -\frac{T}{12 \pi } \sum _i n_i \left( \tilde{m_i} ^{3} - m_i^3  \right) \;,\end{equation}
% where the thermally screened masses are defined in the usual way, $\tilde{m}_i^2=m_i^2 + \Pi T^2$.

\subsection{First order phase transition}

For $a_6 \gtrsim m_{h0}^2/(2 v^4)$ the quadratic and quartic terms in the zero temperature potential \cref{V_phys} change sign, and the potential has a minimum at the origin and a minimum at finite field value with a barrier in between.  The barrier cannot be too large, otherwise the Higgs field will be stuck in the false vacuum until after big bang nuclear synthesis; in fact, for $a_6 \gtrsim 3m_{h0}^2/(4 v^4)$ the minimum at the origin is the true minimum.

A strong first order electroweak phase transition can happen if the potential parameters are close to the critical point of the sign flip
\be
a_6 \sim \frac{m_{h0}^2}{2 v^4} \approx (685 {\rm GeV})^{-2},
\label{demand}
\ee
such that with quantum and temperature effects included there is a
barrier at electroweak scale temperatures.  For small
$a_6 \ll m_{h0}^2/v^4$ the potential is far from the critical point,
and order one loop CW and/or thermal corrections are needed to get a
barrier, which would make the theory non-perturbative.
Note that we also can write \cref{demand} as
\be
-a_4 \sim a_6 v^2 \sim \frac{m_{h0}^2}{2 v^2} \approx 0.12.
\label{demand2}
\ee
In the EW minimum (at zero temperature) the dimension 4 and dimension 6 terms of the potential are balanced. The right-most numerical expression is valid for the measured SM quantities.

%%%%%%%%%%%%%%%%%%%%%%%%%%%%%%%%%%%%%%%%%%%%%%%
\subsection{Temperature corrections}

The thermal corrections to the potential are given in \cref{VT} and we
will briefly show that the effect of the dimension six operator on the
thermal corrections to the potential is small.  Consider a SM
extension with a heavy scalar field $\Phi$ with mass
$m_\Phi^2 \gg T^2$. Its thermal corrections are Boltzmann suppressed
$V_T^{(\Phi)}\propto \e^{-m_\Phi/T}$, see \cref{J_lim}. The heavy degree of
freedom decouples and can be integrated out. In SMEFT the dimension 6
(and higher) operators correct the Higgs mass.  As the thermal
correction depends on the masses, there is thus still
an effect. However, this correction is power law suppressed -- it does
not have the Boltzmann suppression factor, and it corresponds to a
two-loop effect in the UV theory. In a perturbative theory and in the
decoupling limit, it is small.

To be explicit, the Higgs loop gives a contribution to the thermal
potential $V_T^{(h)}\propto T^4J_B(\frac{m_{h}^2}{T^2}) \sim m_{h}^2 T^2 +
m_{h}^3 T +{\cal O}(\frac{m_{h}^4}{T^4})$.  The Higgs mass in the EFT is
\be
m_h^2 = a_2 +3a_4 h^2+5 a_6  h^4
=-\frac12 m_{h0}^2 + a_6 v^4 + \frac32\( \frac{m_{h0}^2}{ v^2} - 4 a_6 v^2\) h^2 + 5 a_6 h^4
\ee
For $a_6 \sim m_{h0}^2/(4v^4)$ the dimension 6 operator can give an order one correction to $h^2$-term, and consequently to the Higgs contribution to the thermal mass and cubic term in the effective potential.  Nonetheless, this will only have a small impact on the total thermal corrections, which are dominated by the loops of the gauge bosons and top quark, as the couplings are much larger than the Higgs self coupling $g^2,g'^2,y^2_t \gg m_h^2/(2v^2) $.  In the SMEFT, the thermal corrections to the potential are thus to a good approximation the same as in the SM.

The SM does not have a strongly first order electroweak phase transition. Let us ignore the dimension six term for the moment and see how this comes about \cite{Cline:2006ts}. Including thermal corrections, the potential in the high temperature limit is of the form
\begin{equation}
  V_{\rm SM} =\frac12 a_2(T) h^2 - \frac{1}{2\sqrt{2}} E\,T h^3+\frac{1}{4} a_4 h^4 \ ,
\end{equation}
with $a_2(T)$ the quadratic term including thermal corrections, and
$E$ the coefficient of the cubic thermal corrections of the bosonic
fields.  At the critical temperature, $T_c$, when the potential has
two degenerate minima at field values $h =0$ and $h=v_c$ (determined
by the conditions $V|_{h=0}=V|_{h=v_c}$ and $\partial_hV|_{h=v_c}=0$), one finds
\begin{equation}
  \label{Rc}
R_c\equiv  \frac{v_c}{T_c}=\frac{E}{\sqrt{2}a_4}=\frac{\sqrt{2} v^2 E}{m_{h0}^2}.
\end{equation}
A strong first order electroweak phase transition requires $
R_N\sim R_c \gtrsim 1$ larger than unity, with $R_N = \frac{v_N}{T_N}$ the
ratio of field value and temperature at the nucleation time when the
phase transition proceeds. This is not realized in the SM, as the
value of $E$ is too small. Adding the dimension 6 term will
predominantly affect the denominator in \cref{Rc}. We can get an
estimate of the strength of the phase transition in SMEFT by using
equation \cref{V_phys} to modify the quartic coefficient  $a_4$ in the denominator 
\begin{equation}
  R_c \approx \frac{E}{\sqrt{2}(m_{h0}^2/(2v^2) - 2 a_6 v^2)}.
  \label{Rc2}
\end{equation}
It is clear that it is now possible to tune the denominator small to get $R_c >1$ if we take $a_6$ close to the critical value in \cref{demand}.  This is basically the derivation of the arguments in the previous subsection.

%%%%%%%%%%%%%%%%%%%%%%%%%%%%%%%%%%%%%%%%%%%%%%%%%%%%
\subsection{Validity of the SMEFT}

 \begin{figure}[t]
	\centering
	\includegraphics[width=0.45\textwidth]{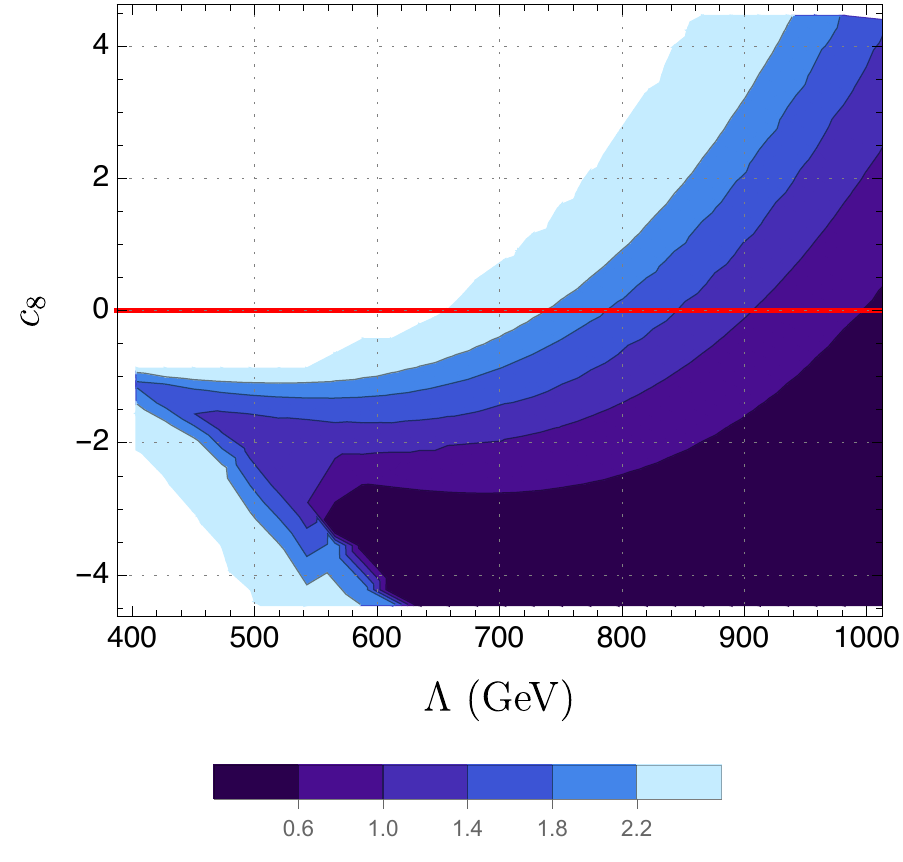} \hspace{0.5cm}
        \includegraphics[width=0.45\textwidth]{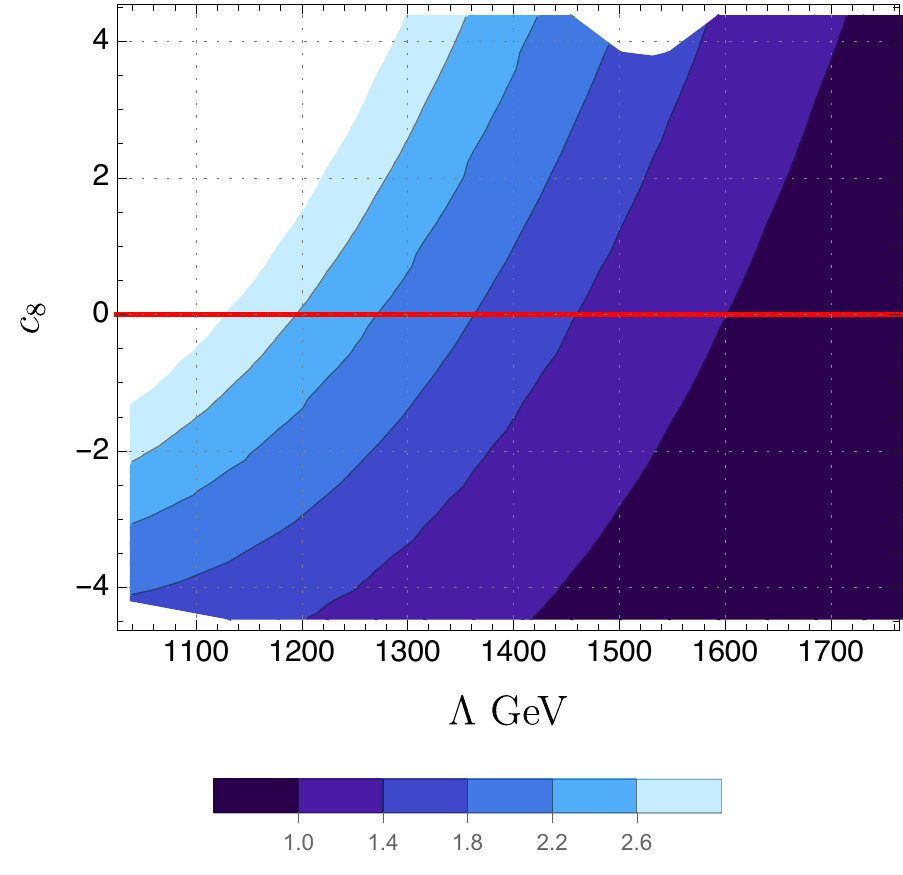} \hspace{0.5cm}
	\caption{\it The strength of the first order phase transtion
          $R_N =\frac{v_N}{T_N}$ as function of $\Lambda$ and $c_8$ in
          SMEFT (left plot) and for a dark Higgs mass of $m_h=80$\,GeV
          and all other parameters as in the SM (right plot). The red
          line corresponds to the SMEFT truncated at dimension 6,
          which is a good approximation only when the dimension 8
          operator is negligible.  In the white region the phase transition is 2nd
          order.  }\label{F:EFT_PT}
\end{figure}

The first demand on the validity of the EFT is that the dimension 6 corrections to the kinetic terms are small
\be
2c_{\rm kin} v^2 <\frac12
\label{demand0}
\ee
This assures the dimension eight and higher order derivative operators
have a subdominant effect. This condition was already used in defining
the canonical field to arrive at \cref{Va}.
%; if it were not satisfied the potential for the canonical field would actually no longer be a polynomial in $h^2$.

We further demand that the higher order corrections to the potential are small as well, starting with the dimension 8 term. We parameterize the potential including dimension 8 terms as
\be
V= \frac12 a_2 h^2 +\frac14 a_4 h^4 + \frac16 a_6 h^6+\frac18 a_8 h^8 =
 \frac12 a_2 h^2 +\frac14 a_4 h^4 +
 \frac16\frac{h^6}{\Lambda^2}+\frac{c_8}{8} \frac{h^8}{\Lambda^4}
 \label{V_SMEFT}
\ee
with the cutoff scale $\Lambda^2=a_6^{-1}$ and $c_8 = a_8/a_6^2$.
In the EW vacuum the dimension 4 and dimension 6 operators are balanced \cref{demand2}; requiring the dimension 8 contribution to be small thus leads to the condition
\be
\frac{|a_8| v^2}{ a_6}  = \frac{c_8 v^2}{\Lambda^2}< \frac12
\label{valid1}
\ee

The conditions \cref{demand0,valid1} assure that the separation of
scales between the light and heavy degrees of freedom required for the
EFT expansion in operators of increasing mass dimensions converges.
Depending on the accuracy aimed for, the EFT can than be truncated at
a given mass dimension.  To get accurate results for the parameters of
the phase transition, e.g. the nucleation temperature and the strength
of the phase transition $R_N=\frac{v_N}{T_N}$,  in SMEFT at
dimension 6 operators is only possible if the
ratio in \cref{valid1} is sufficiently small. In the left plot of
\cref{F:EFT_PT} we show the numerical results for $R_N$ as a function
of the cutoff scale $\Lambda$ and coefficient of the dimension 8
operator $c_8$. All parameter scans are done with the
\texttt{CosmoTransitions} package \cite{Wainwright:2011kj}; details on
our implementation can be found in \cref{A:scan}.

Turning on the dimension 8 operator and changing $c_8$ from zero to order one values generically gives a change in $R_N$ of 10\% or larger. From this we conclude that a quantitative description of the phase transition -- with parameters determined within 10\% accuracy -- requires
\be
|c_8| \lesssim 1
\label{valid2}
\ee
which is a stronger condition than convergence of the EFT
\cref{valid1}. Although this conditions seems independent of the scale
separation in \cref{valid1}, this is not the case, as will become
clear as we discuss dark sector phase transitions in the next
subsection.  We further note that \cref{demand0,valid1} assures the
EFT validity in the electroweak vacuum, which not necessary implies
the same during the phase transition (although it gives a good
indication).  Condition \cref{valid2}, on the other hand, derives
directly from the phase transition dynamics.

%%%%%%%%%%%%%%%%%%%%%%%%%%%%%%%%%%%%%%%%%%%%%%%%%%%%
\subsection{Dark  phase transition}
\label{s:DS}

Let us also discuss the validity of an EFT description for a strongly
first order phase transiton (SFO-PT) in a dark sector with a potential
$V(|H_D|^2)$ for the dark Higgs field $H_D$ that mimics that of the
SM. The important difference is that the dark Higgs mass and vev, as well as the thermal corrections (in particular the size of the cubic term $E$) are now all free parameters not fixed by experiment. We will assume that the dimension 6 terms are essential for obtaining a strong first order phase transition, and that in its absence $R_N < 1$; otherwise the EFT description can always be made to work by taking the cutoff scale arbitrarily large. This leads to the condition
\be
\frac {m_{h_D0}^2}{v_D^2} \gtrsim \sqrt{2}E \sim 0.03 n_g g^2
\label{dim6_need}
\ee
if the dark Higgs couples to $n_g$ thermal bosonic degrees of freedom with coupling $g$.

The requirement on the parameter space for a SFO-PT can be read off
again from the expression for $R_c$ in \cref{Rc2}. As we have seen, in
the SM the measured Higgs mass is far off the critical value, and the
correction of the dimension 6 term to the potential needs to be order
one. In the dark sector the ratio $ m_{h_D0}^2/(2v_D^2)$ can be small and
much closer to the critical value, allowing the dimension 6 term to
only give a small (but essential) correction. This allows for a larger
cutoff scale and a thus a better convergence of the EFT expansion.

This is confirmed by our numerical results. The right plot in
\cref{F:EFT_PT} shows $R_N$ as a function of the cutof $\Lambda$ and
dimension 8 coefficient $c_8$ for a dark Higgs mass of
$m_{h_D0} =80\,$GeV; the dark Higgs vev and thermal spectrum are chosen
as in the SM.  We indeed see that for a smaller ratio
$ m_{h_D0}^2/(2v_D^2)$ than in the SM, the cutoff scale of the dimension 6
operator is much larger in the parameter space for a SFO-PT. As a
consequence of the much larger separation of scales, the dependence on
$c_8$ is less as well.

%%%%%%%%%%%%%%%%%%%%%%%%%%%%%%%%%%%%%%%%%%%%%%%%%%%%
\section{Matching the SMEFT to UV theories}\label{sec:matching}

We consider the SM augmented with heavy degrees of freedom. If the
heavy fields are flavor diagonal, the low energy EFT can be matched to
the UV theory using the covariant derivative expansion method of
\cite{Gaillard:1985uh,Cheyette:1987qz,delAguila:2016zcb,Henning:2014wua}
(for a SMEFT review, see \cite{Brivio:2017vri}). For flavor
off-diagonal new physics the more general SMEFT structure of \cite{Drozd:2015rsp,Henning:2016lyp,Ellis:2017jns,Fuentes-Martin:2016uol} can be used, which also includes the effects of mixed diagrams with both heavy and light fields in the loop.

We focus on the simplest possibility of adding a single (multiplet)
field to the SM, and use the covariant derivative expansion method. We
will briefly comment on more complicated set-ups with multiple heavy
fields in \cref{sec:multifield}. To match the UV theory onto SMEFT,
the effective action is calculated and expanded in powers of the mass
parameter of the heavy field.  We will discuss tree level and loop
level matching in turn.

%%%%%%%%%%%%%%%%%%%%%%%%%%%%%%%%%%%%%%%%%%%%%%%%%%%
%%%%%%%%%%%%%%%%%%%%%%%%%%%%%%%%%%%%%%%%%%%%%%%%%%%
%%%%%%%%%%%%%%%%%%%%%%%%%%%%%%%%%%%%%%%%%%%%%%%%%%%

\subsection{Tree-level matching}

The SMEFT higher dimensional operators can be generated by tree level diagrams if the heavy field $\Phi$ has a non-zero vacuum expectation value (vev). This limits the possibilities to (effective) scalar fields. Furthermore, the model space is severely limited by electroweak precision constraints. Specifically, if we add new scalar degrees of freedom with non-zero vev $v_i$ to the SM, the $\rho$-parameter becomes
\be
\rho = \frac{\sum_i (4 I_i(I_i+1)-Y_i^2)v_i^2}{\sum_i Y_i^2 v_i^2},
\ee
with $I_i,Y_i$ the isospin and hypercharge of the additional multiplets.  For singlets with $I=Y=0$ or additional doublets with $I=1/2, Y=1$ the $\rho$-parameter is the same as in the SM $\rho=\rho_{\rm SM} $.  For all other multiplets $X$ the precise measurement of $\rho-\rho_{\rm SM} =0.0038 \pm 0.00020$ \cite{10.1093/ptep/ptaa104} severely limits the size of the vev $v_X^2/v^2 \lesssim 10^{-2}$.  This implies that the mass of this multiplet has to be in the $10\,$TeV range or higher, and the set-up is not interesting for the EWPT. This leaves Higgs-singlet extensions and two Higgs doublet models as the interesting cases for tree level matching, which we will discuss in detail in the next section. Here we briefly recap the generic tree-level matching results.

Write the UV Lagrangian for the heavy complex scalar (mulitplet) $\Phi$ in the form
\be
{\cal L}_{\rm UV} = (\Phi^\dagger B +{\rm h.c.})+ \Phi^\dagger
(-D^2 -M^2 - U) \Phi + {\cal O}(\Phi^3),
\label{generalL}
\ee
with $M$ the field-independent mass of the heavy field, and $B(H),U(H)$ parameterizing the coupling to the Higgs field. Since we are only interested in the Higgs dynamics we can replace covariant derivatives with partial derivatives $D^2=\partial^2$.  For a real scalar field we can substitute $\Phi =\Phi^\dagger \to \phi/\sqrt{2}$.  The scalar vev will be non-zero for $B \neq 0$, and we can integrate it out using its equation of motion:
\be (P^2-M^2-U) \Phi=-B +{\cal O} (\Phi^2)
\ee
with $P^2 = -\partial^2$.  To leading approximation (for small
couplings of the ${\cal O}(\Phi^3)$ terms) the solution is
\be
\Phi_c =- \frac{1}{P^2-M^2-U} B.
% = \frac1{M^2}B + \frac1{M^4}(P^2-U)B +\frac{1}{M^6} (P^2-U)^2 B+..
\label{Phi_c}
\ee
We can improve this perturbatively by replacing $B \to B+ {\cal
  O}(\Phi^2_c)$, where the higher order terms are evaluated at the 0th
order solution $\Phi_c$.  Plugging back in the action gives
\begin{align}
  {\cal L}_{\rm tree-level} &= - B^\dagger \frac{1}{P^2-M^2-U} B 
                      + {\cal O}(\Phi_c^3) \nn \\
                      &= \frac1{M^2} B^\dagger B+ \frac1{M^4} B^\dagger (P^2-U)B+ \frac1{M^6} B^\dagger (P^2-U)^2 B+{\cal O}(M^{-8} )+ {\cal O}(\Phi_c^3),
                        \label{LEFT_tree}
\end{align}
where we expanded in large $M^2$ and demand that EFT expansion is valid (cf. \cref{demand0,valid1})
\begin{equation}
  \frac{P^2-U}{M^2}\ll 1, \qquad \frac{|B|^2}{M^2} \ll v^2.
  \label{validity}
\end{equation}
%

%%%%%%%%%%%%%%%%%%%%%%%%%%%%%%%%%%%%%%%%%%%%%%%%%%%%
%%%%%%%%%%%%%%%%%%%%%%%%%%%%%%%%%%%%%%%%%%%%%%%%%%%%
%%%%%%%%%%%%%%%%%%%%%%%%%%%%%%%%%%%%%%%%%%%%%%%%%%%%

\subsection{Loop-level matching}

In addition to the tree level matching there will also be loop level matching contributions to the SMEFT operators. Focus on a scalar extension again with the UV Lagrangian given in \cref{generalL}. For theories with $B=0$ there are only loop-level matching contributions. Vanishing of $B$ may be enforced by symmetries, for example, in Higgs-singlet extensions the linear $B$-term can be forbidden by a $Z_2$ symmetry under which the singlet transforms as $\Phi \to -\Phi$. 

Calculating the one-loop corrections to the effective action and expanding in powers of  $M^2$, the results can be matched to the SMEFT Lagrangian \cite{Henning:2014wua}.
The logarithmically divergent $M^4$ and $M^2,M^0$-terms contribute to the cosmological constant, and the $\mu$-term and Higgs self-coupling respectively.  The quadratic and quartic counterterms in the SMEFT are fixed by our on shell renormalization condition. Let's then focus on the finite dimension 6 and dimension 8 operators
\begin{align}
  (4\pi)^2 c_i^{-1} {\cal L}_{\rm loop}&=\tr \[\frac1{M^2}\( -\frac1{6} U^3+\frac1{12}(\partial_\mu U)^2\)
                                         +        \frac1{M^4}\( \frac1{24} U^4-\frac1{12} U(\partial_\mu U)^2+\frac1{120} (\partial^2 U)^2\)\]
                                         %\nn \\
                                         % &+   \frac1{M^6}\( -\frac1{60} U^5+\frac1{20} U^2(\partial_\mu U)^2+\frac1{30} (U\partial_\mu U)^2\) + {\cal O}(M^{-8})
\label{LEFT_loop}                                             
\end{align}
with $c_i=1/2 \, (1)$ for a real (complex) scalar. The trace over the gauge indices depends on the representation of the heavy d.o.f. The results \cref{LEFT_loop} can also be applied to UV theories with a heavy fermion or gauge field, with $c_i$ and $U$ taken as in \cite{Henning:2014wua}.  For fermions the loop contributions have a minus sign, and are of no help for obtaining a SFO-PT. If the Higgs field is charged under a gauge symmetry that is broken at a large scale, there can be a $g^2 h^2 B^2$-interaction term in the Lagrangian, with $B$ the heavy gauge field, and the loop matching results will qualitatively be similar to the scalar field case with a $\kappa h^2 |\Phi|^2$-interaction \cite{Henning:2014wua}.

The  result \cref{LEFT_loop}  is valid if both the EFT expansion in powers of $M^{-2}$ and the perturbative loop expansion holds, which gives respectively
\be
\frac{U}{M^2}\ll 1, \quad \& \quad \frac{c_i  \kappa^2}{(4\pi)^2} \ll 1.
\label{valid}
\ee
%

%%%%%%%%%%%%%%%%%%%%%%%%%%%%%%%%%%%%%%%%%%%%%%%%%%%%%
\section{Phase transitions in UV theories and in SMEFT }\label{sec:explicitmatch}

In this section we analyse the parameter space for a strongly first order electroweak phase transitions in specific UV theories, and compare that with the results obtained using the EFT description. We focus on the most interesting cases of singlet extensions of the SM, both with tree-level and only loop level matching, and two Higgs doublet models. We will also discuss how results can be adapted to dark phase transitions.

\subsection{Loop level matching: scalar extensions}
\label{s:looplevel}

Consider a $Z_2$-symmetric scalar extension of the SM, with no linear interaction in \cref{generalL} and $B=0$. The SMEFT operators then only arise from the loop diagrams as in \cref{LEFT_loop}. For a singlet field the interaction term is
\be
U = \frac12 \kappa  h^2
\label{Uloop}
\ee
expressed in unitary gauge $|H|^2 =h^2/2$.  For scalar multiplets there will be gauge generators as well and the dimension 6 operator in \cref{LEFT_loop} may be enhanced by trace factors. However, generically the dimension 8 and higher terms will then be likewise enhanced, and the perturbativity constraint is stronger for larger $c_i$ -- we thus do not expect that the multiplet structure will significantly improve the EFT validity \cref{valid} and for simplicity we work with \cref{Uloop} above.

We can then read off the explicit Wilson coefficients by comparing the general expression in \cref{LEFT_loop} with \cref{L_wilson}
\be
c_H = -\frac{c_i}{6M^2(4\pi^2)} \kappa^3 ,\quad c_{\rm kin} = \frac{c_i}{12M^2(4\pi)^2} \kappa^2
\quad \Rightarrow\quad a_6 = \frac{c_i}{M^2(4\pi)^2} \big(\frac18\kappa^3 +\frac16 \kappa^2 \lambda\big).
\ee
The requirement that $c_{\rm kin}$ gives a small correction to the kinetic term \cref{demand0} is satisfied automatically for a perturbatively small coupling $\kappa$.  For the singlet to have an impact on the phase transition dynamics, and thus for the dimension 6 term in the SMEFT approximation to be sufficiently large, requires relatively large couplings  $\kappa_i \gtrsim 1$.  As a first approximation we can then neglect $c_{\rm kin}$ and thus the $(\partial U)$ derivative terms in \cref{LEFT_loop}.

A strong first order electroweak phase transition requires balancing the dimension 4 and dimension 6 terms \cref{demand2}. Neglecting the derivative terms this gives
\be
\frac14 \frac{ c_i \kappa^2}{(4\pi)^2} \frac{U}{M^2} \sim \frac {m_{h0}^2}{v^2}  \approx 0.12.
\ee
This cannot be satisfied without either violating the EFT expansion or the loop expansion  \cref{valid}.  We thus conclude that the SMEFT framework with only loop-suppressed higher order operators cannot be used for SFO-EWPTs.

\subsubsection{Dark sector}
Turning our attention to dark sectors, we recall that for a dark Higgs potential $V(|H_D|^2)$, with $H_D$ the dark sector Higgs field, a strong first order phase transition is possible if 
\be
\frac{m_{h_D 0}^2}{ v_D^2} \sim \frac12 \frac{ c_i \kappa^2}{(4\pi)^2} \frac{U}{M^2} 
= \frac{\kappa }{4} \frac{ c_i \kappa^2}{(4\pi)^2} \frac{v_D^2}{M^2} \ll 1
\ee
which can be satisfied for sufficiently small dark Higgs mass.
Note, however, that the $({m_{h_D 0}^2}/{ v_D^2} )$-ratio can also not be too small if the dimension 6 term is to be essential for the SFO-PT, see \cref{dim6_need}, which limits the applicability of the EFT framework for loop-level matching.

%%%%%%%%%%%%%%%%%%%%%%%%%%%%%%%%%%%%%%%%%%%%%%%%%%%%%%%%%%%%%%%%
\subsection{Tree level matching: two Higgs doublet models}

In two Higgs doublet models (2HDMs) both Higgs fields can obtain a vev. If there is a separation of scales between the Standard Model-like Higgs and the heavy Higgs, this allows for an EFT description with tree level matching.  The most general effective potential in the 2HDM is given by \cite{Lee:1973iz,Gunion:1989we,Branco:2011iw}
\begin{align}
  V_H& = m_{11}^2 \Phi_1^\dagger \Phi_1 + m_{22}^2 \Phi_2^\dagger \Phi_2 -(m_{12}^2 \Phi_1^\dagger \Phi_2 +{\rm h.c.}) \nn \\
  &+\frac12 \lambda_1 (\Phi_1^\dagger \Phi_1)^2 +\frac12 \lambda_2 (\Phi_2^\dagger \Phi_2)^2
  + \lambda_3 (\Phi_1^\dagger \Phi_1) (\Phi_2^\dagger \Phi_2)
  +\lambda_4 (\Phi_1^\dagger \Phi_2) (\Phi_2^\dagger \Phi_1) \nn \\
 & +\[ \frac12 \lambda_5 (\Phi_1^\dagger \Phi_2)^2
  +\lambda_6 (\Phi_1^\dagger \Phi_1) (\Phi_1^\dagger \Phi_2)
  +\lambda_7 (\Phi_2^\dagger \Phi_2) (\Phi_1^\dagger \Phi_2) +{\rm h.c.})  \].
\end{align}
%
% From the 6 real plus 4 complex parameters (in general $m_{12}^2$ and $\lambda_{5,6,7}$ can be complex), 3 are not independent and can be rotated away leaving 11 free parameters.
We identify $\Phi_1=H$ with the Standard Model-like Higgs and $\Phi_2=\Phi$ with the heavy degree of freedom.  Then
\be
B = A_2 H +A_0 H (H^\dagger H),
\quad U = \lambda_3 H^\dagger H + \lambda_4 H H^\dagger,
\ee
with $A_2 =-m_{12}^2$, $A_0 = -\lambda_6$ (the subscript on $A$ denotes the mass dimension of the coupling), and $M^2 =m_{22}^2$.

The leading corrections to the kinetic terms are
\be
{\cal L}_M \supset  -\frac1{M^4}\[A_2 H^\dagger +A_0 (H^\dagger H) H^\dagger \]\Box\[ A_2 H +A_0 H (H^\dagger H)\],
\ee
which are perturbatively small \cref{demand0} for
\be
\frac{|B|^2}{M^4} \ll v^2 \quad \Rightarrow \quad
\frac{|A_2|}{M^4} \ll 1 \quad \& \quad \frac{|A_0|^2 v^4}{M^4} \ll 1.
\label{kin_pert}
\ee
The $\Box$-corrections above are subdominant for $\lambda_{3,4} > \lambda$, which follows from using the Higgs equations of motion $\Box h = -\partial_h V_{\rm SM}+{\cal O}(M^{-2})$. This is indeed the limit of interest for a SFO-EWPT, which requires a strong coupling between the Higgs and the heavy field.  Ignoring then the dimension 6 derivative operators the effective potential is
\begin{align}
  V_{\rm EFT} = \frac{h^6}{8M^2}\( -|A_0|^2 +\frac{2\kappa {\rm Re}(A_0 A_2)}{M^2} - \frac{\kappa ^2 |A_2|^2}{M^4}\),
\end{align}
with $\kappa = \lambda_3+\lambda_4$. The dimension 6 term is negative,  which does not work for obtaining a SFO-EWPT in SMEFT at this mass dimension.\footnote {It is also not an option to take $M^2<0$ to reverse the sign, as this leads to an  unstable  electroweak vacuum with a tachyonic heavy mass eigenstate. Indeed, the requirement the sum of the mass eigenvalues is positive cannot be obtained with a valid EFT expansion $U/M^2 \ll 1$ (see \cref{valid} and \cref{kin_pert}).} An electroweak minimum separated by a barrier from the mininum at the origin can only be obtained balancing the dimension 6 with positive dimension 8 terms. This clearly violates the EFT expansion, and we conclude that the SMEFT framework fails to describe SFO-EWPT in two Higgs doublet models.

\subsection{Tree level matching: Higgs-singlet models}

Consider the SM coupled to a real singlet field $s$ with Lagrangian
\be
{\cal L} \supset |D H|^2 +\frac12 (\partial s)^2 -\( \mu_0^2 |H|^2+ \frac12 M^2 s^2 +\lambda_0|H|^4 +\frac{\lambda_s}4 s^4 
+\frac{\kappa}{2}|H|^2 s^2-A_1 |H|^2 s+ \frac{g_s A_1}{3} s^3\)
\label{L_S} .
\ee
Setting $A_1$ to zero, the Lagrangian has a discrete $Z_2$-symmetry. The tree level matching contributions are thus proportional to $A_1$. Specifically, we identify (cf. \cref{generalL})
\be
U = \kappa |H|^2 = \frac12 \kappa h^2, \quad
B= \sqrt{2}A_1 |H|^2 = \frac1{\sqrt{2}} A_1 h^2
\label{UB_S}
\ee
The leading SMEFT Lagrangian operators are\footnote{We assumed $g_s, \lambda_s \ll \kappa$. Since $|A_1|/M^2$ can be order one these corrections need not be small.  The $g_s$-corrections can be included replacing $\kappa \to \kappa + \frac23 g_s^2\frac{|A_1|^2}{M^2}$ in $a_6$, and $\kappa \to \kappa + \frac2 g_s^2\frac{|A_1|^2}{M^2}$ in $a_8$. This  however does not change the qualitative analysis of the 1PT as we can simply  replace $\kappa \to {\rm max}(\kappa, g_s^2\frac{|A_1|^2}{M^2})$ in \cref{val_eq1,val_eq2}.}
\begin{align}
   {\cal L}_{\rm EFT} &\supset -\frac{|A_1|^2}{8M^4}  h^2\Box  h^2
                -
 \frac{ \kappa|A_1|^2}{16M^4}h^6
                        + \frac{|A_1|^2 \kappa^2 }{32M^6} h^8+...
  %                      \nn \\
  % &=\frac12 (\partial h)^2 
  %               -\[\frac12 a_2  h^2 +\frac14 a_4 h^4+
  %                 \frac16 a_6 h^6+ \frac18 a_8h^8+...\]
      \label{singlet_smeft}
\end{align}
The dimension 6 $\Box$-operator that corrects the Higgs kinetic term is small enough for the EFT expansion to be valid \cref{validity}, and singlet loop diagrams constructed with the $A_1$-coupling are pertubative, for respectively
\be
\frac{|A_1|^2 v^2}{4M^4} \ll 1 \quad \& \quad
\frac{|A_1|^2}{M^2} \ll (4\pi )^2
\label{A1_bound}
\ee
which are both not very strong constraints. 

The matching is performed at the heavy scale $M$. We will neglect the running of the parameters between this scale and the $Z$-scale relevant for the electroweak phase transition, as the separation is not large and we expect this effect to be small.  Using $- h^2\Box h^2 =4 h^2(\partial h)^2$ we read off $c_{\rm kin} = |A_1|^2/(2M^4)$ and $c_H = -\kappa |A_1|^2/(2M^4)$. Comparing with \cref{akin} this gives a SMEFT effective potential $V = \sum_n \frac{1}{2n}a_{2n} h^{2n}$ for the canonically normalized Higgs field, with
\be
% \lambda = \lambda_0 -\frac{|A_1|^2}{4M^2} ,
a_2 =\mu^2,\quad
a_4 =  \lambda-\frac23 \frac{|A_1|^2 \mu^2}{M^4}, %= \lambda_0 -\frac{19}{12} \frac{A^2}{M^2},
\quad
a_6 =  \frac38 \frac{|A_1|^2}{M^4}\(\kappa-\frac83 \lambda\),
\quad
a_8 =-\frac{|A_1|^2 \kappa^2}{4M^6} +{\cal O}\big(\lambda, \frac{\mu^2}{M^2} \big)\ .
\label{mapping}
\ee
For simplicity, in this procedure we have neglected the dimension 8 derivative operators,
and the corrections to $a_8$ from rewriting the potential in terms of the canonical field. 
In the parameter space of interest the potential corrections dominate over the derivative corrections, and this is a good approximation. The parameters $(\mu,\lambda)$ are fixed by the on shell renormalization conditions discussed below \cref{V_phys}.  In addition to these tree level matching results there will also be the subdominant loop level corrections, as discussed in \cref{s:looplevel}. We will neglect them in our discussion below, as they will not change our qualitative results; but for a precise quantitative discussion they should be included as well.

A 1PT can arise if \cref{demand} is satisfied, which gives
\be
 0.12 \simeq \frac{m_{h0}^2}{2v^2} \sim a_6 v^2=  \frac38 \frac{\kappa |A_1|^2
   v^2}{M^4} \ll  \frac34\frac{|A_1|^2 }{M^2},
 \label{val_eq1} 
 \ee
 where in the last step we used $U\ll M^2$, see \cref{validity}, to
 assure the convergence of the EFT expansion.  We used here that
 $\lambda \ll \kappa$ in the parameter space of interest, which is the
 statement that the dimension 6 $\Box$-corrections are subdominant.
 The above condition can be marginally satisfied and SMEFT may
 adequately describe the SFO-EWPT for $|A_1|^2 \sim M^2$ (satisfying
 \cref{A1_bound}) and for $U/M^2 \gtrsim 0.1$ not too small (and thus
 for large $\kappa$). The cutoff scale of the dimension 6 operator and
 the coefficient of the dimension 8 operator, both defined in
 \cref{V_SMEFT} are
 \be
 \Lambda^2 = \frac83 \frac{M^4}{\kappa |A_1|^2}, \quad
 c_8 = - \frac{16}{9} \frac {M^2}{|A_1|^2} .
 \label{val_eq2} 
 \ee
Since $c_8 = {\cal O}(1)$ for $|A_1|^2 \sim M^2$, for an accurate description of the phase transition the dimension 8 terms should be included, as follows from \cref{demand2}.

Finally, we note that there is a phenomenological constraint on the vev of the scalar field denoted by $s_0$. Defining the mixing angle via $h_1 = h\, \cos \theta +s \,\sin \theta$  with $h_1$ the lightest, mostly-Higgs mass eigenstate, gives (the full expression for $y$ is given in \cref{Amixing})
 \be
\tan \theta = \frac{y}{1+\sqrt{1+y^2}}, \qquad y =\frac{2 v(A_1-  \kappa s_0)}{M^2} + {\cal O}(M^{-4}) .
\label{mixing}
 \ee 
For singlet masses $m_s \gtrsim 600\,$GeV the experimental bound was derived in Ref. \cite{Huang:2017jws} to be $\cos \theta \gtrsim 0.97 \,(0.91)$ at $2\sigma \,(3\sigma)$.

 \subsubsection{Numerical results}
 
\begin{figure}
  	\begin{subfigure}{1\textwidth}
	\centering
	\includegraphics[width=0.4\textwidth]{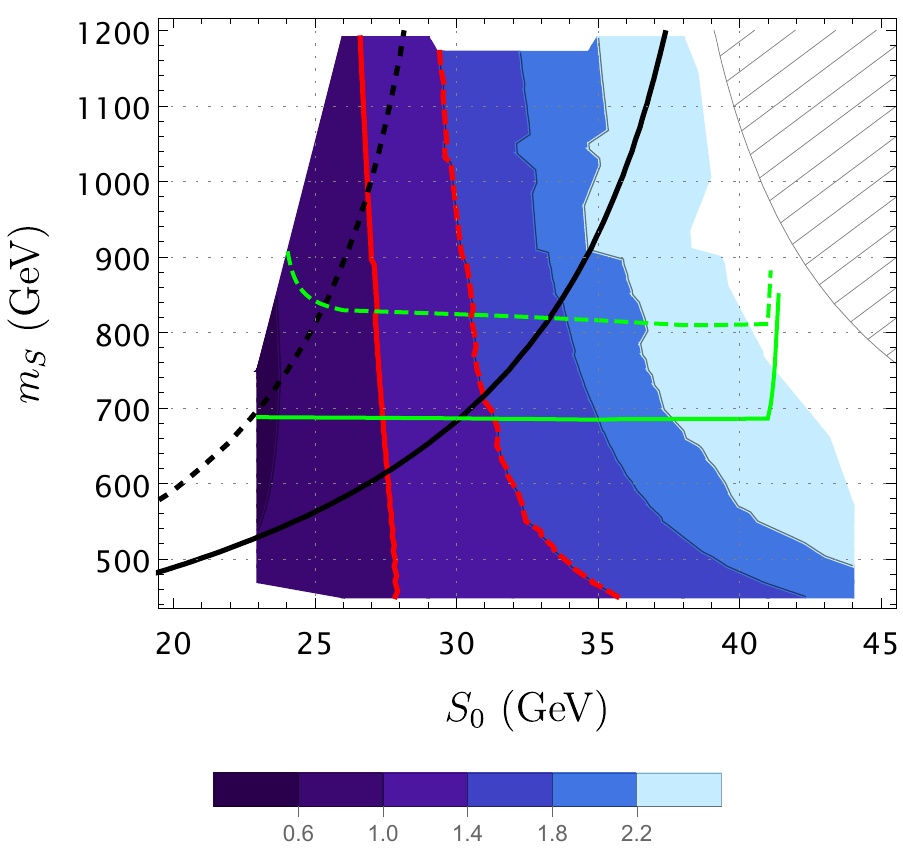} \hspace{0.5cm}
        \includegraphics[width=0.4\textwidth]{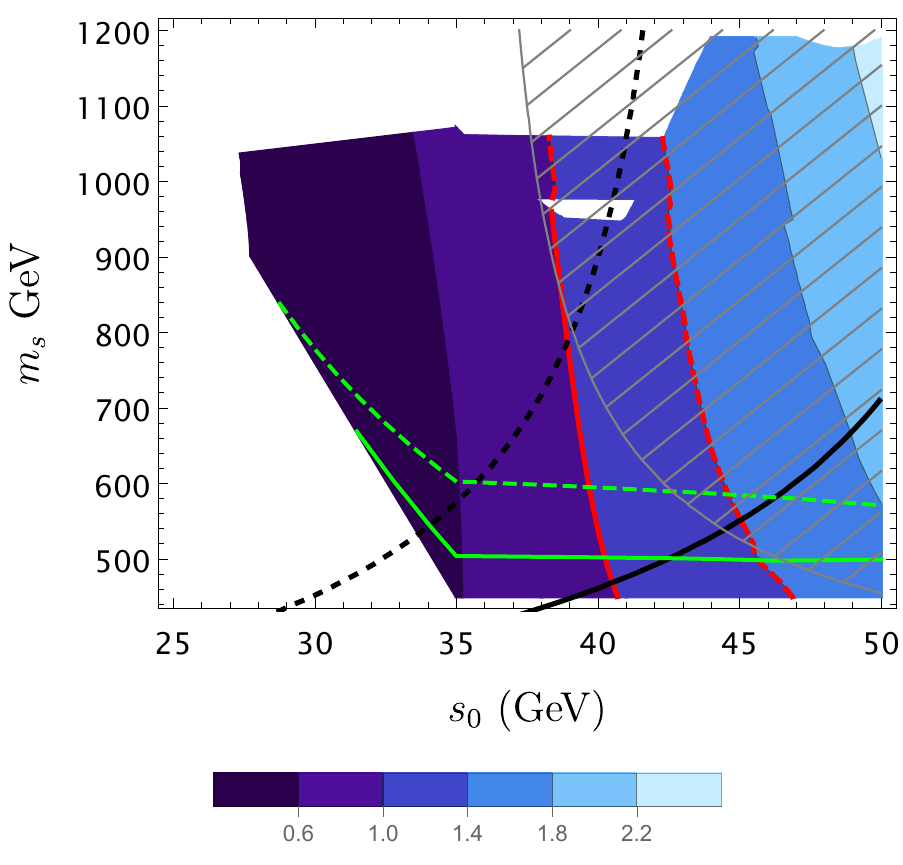}   
\hspace{0.5cm}		\vspace*{-0.2cm}
		\caption{\it Strength of the
          first order phase transition $R_N$ as
          function of the singlet vev $s_0$ and mass $m_s$ in the
          Higgs-singlet model with $\kappa=4$ (left plot) and
          $\kappa=2$ (right plot). The gray arched area in top right
          corner is excluded by the $2\sigma$ constraint on the mixing
          parameter; the $3\sigma$ constraint does not affect parameter space.
   }
          \end{subfigure}
	\begin{subfigure}{1\textwidth}
	\centering
        \includegraphics[width=0.4\textwidth]{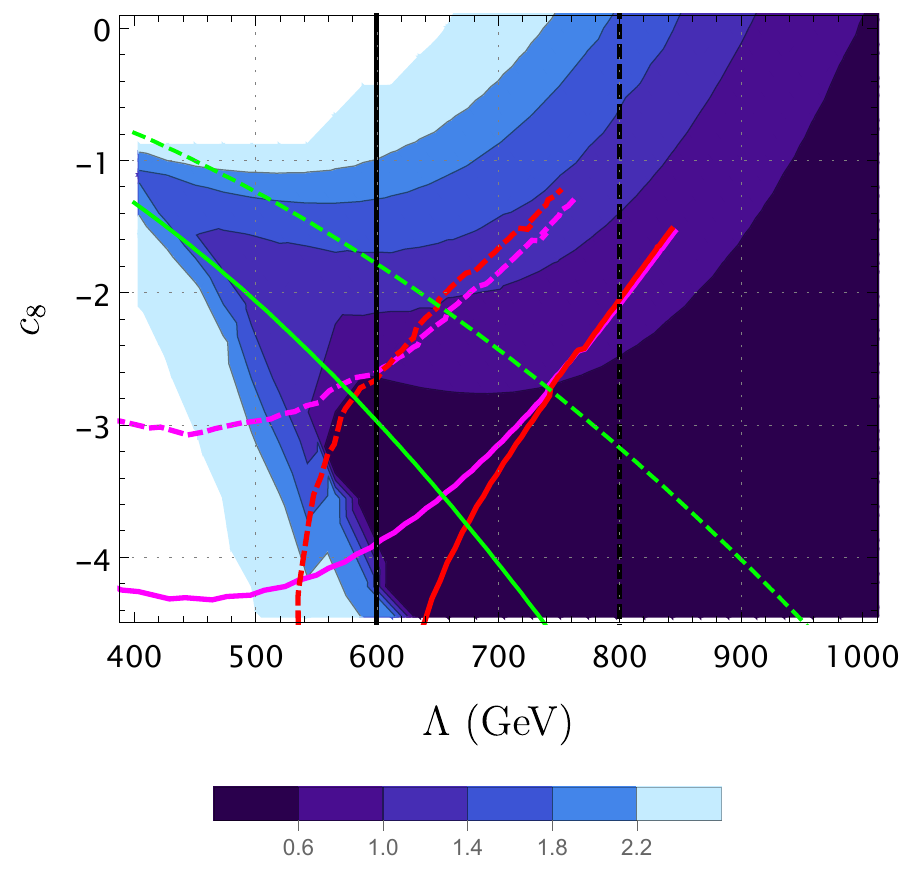}
        \hspace{0.5cm}
        \includegraphics[width=0.4\textwidth]{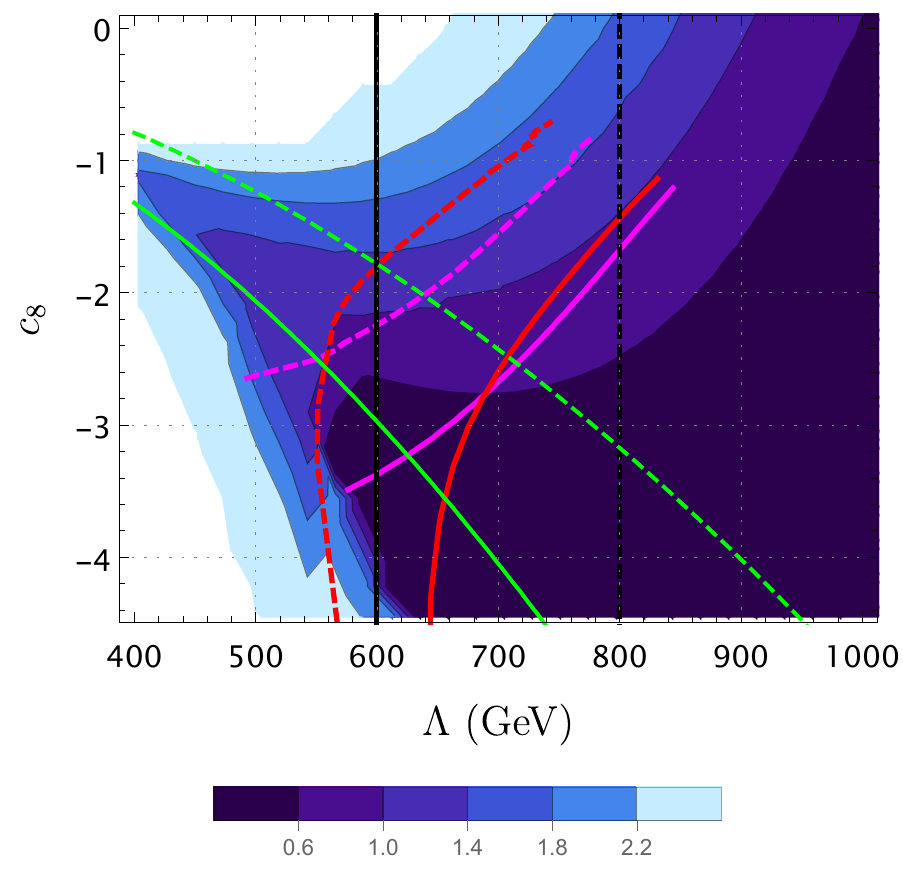}
        \hspace{0.5cm}		\vspace*{-0.2cm}
 	\caption{\it Strength of the
          first order phase transition $R_N$ as
          function of the EFT parameters $\Lambda$ and mass $c_8$.
         The red lines for constant $R_N$ in the singlet model are
         mapped to the corresponding SMEFT parameters
          for $\kappa=4$ (left plot) and
          $\kappa=2$ (right plot). }
      \end{subfigure}
      \caption{\it Parameter space for a SFO-EWPT in the singlet extension of the standard model (top row) and
        SMEFT (bottom row) . The solid (dashed) red lines correspond to
        $R_N =\frac{v_N}{T_N}=1\, (1.4)$ in the singlet model; the solid (dashed)
        green lines to $a_8 v^2/a_6 =0.5\, (0.3)$ and the
      solid (dashed) black lines lines to
        $\Lambda =600\,(800)\,$GeV in SMEFT. These lines are mapped
        between Higgs-singlet model and SMEFT parameters and shown in both plots for
        $\kappa =4$ (right side) and $\kappa =2$ left side; in both
        cases $(g_s,\lambda_s) = (0,1)$.}
      \label{F:singlet}
    \end{figure}

\begin{figure}
\centering
\includegraphics[width=0.4\textwidth]{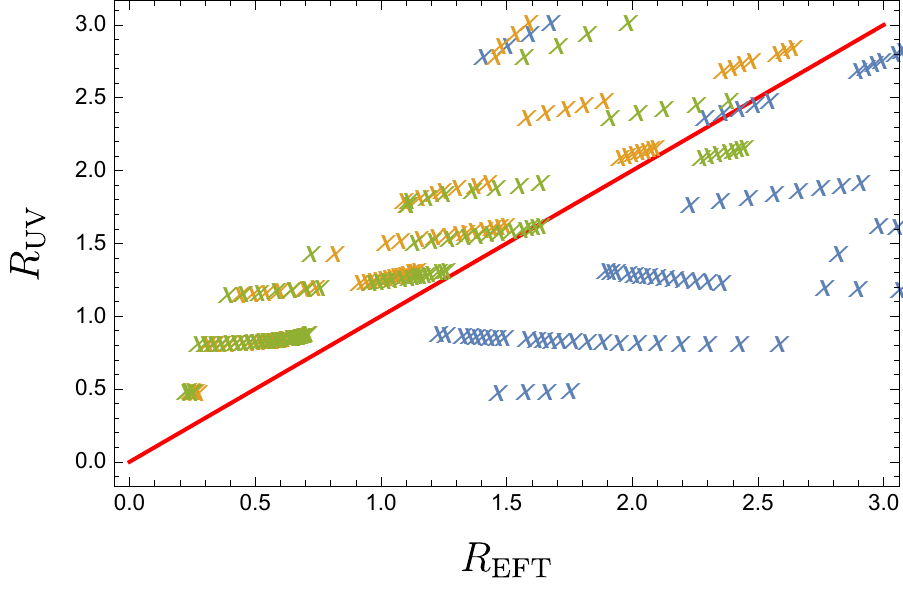} \hspace{0.5cm}
\includegraphics[width=0.4\textwidth]{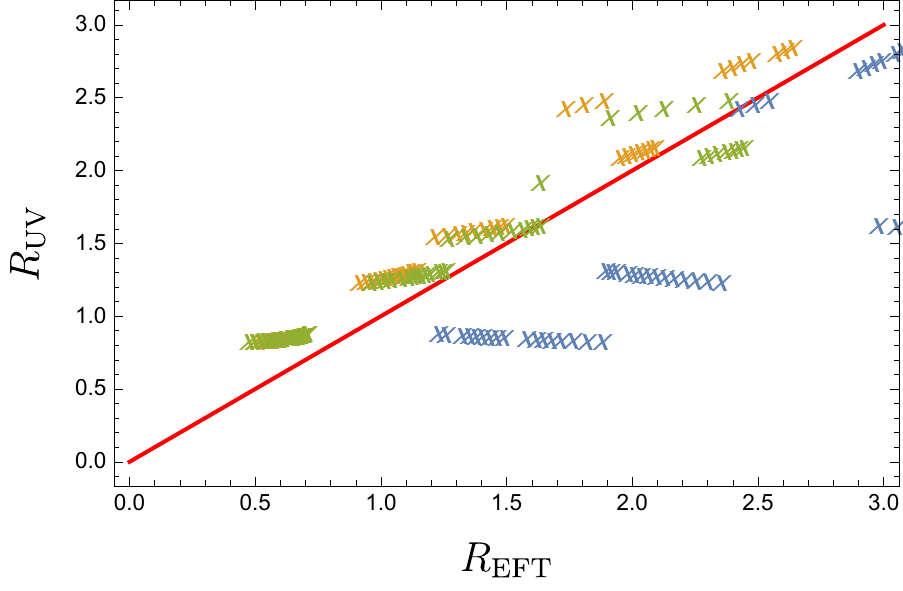}
\caption{\it $(R_N)_{\rm EFT}$ on the x-axis vs. $(R_N)_{\rm UV}$ on the y-axis for points from the singlet scan with $\kappa=4$ in \cref{F:singlet}; in the left plot we included all points with $c_8 v^2/\Lambda^2 <0.5$, and in the right plot all points with $c_8 v^2/\Lambda^2<0.3$. The red line corresponds to $(R_N)_{\rm EFT}=(R_N)_{\rm UV}$. The blue, orange, green points correspond to SMEFT with only dimension 6 operators, SMEFT with dimension 6 and dimension 8 operators, and SMEFT with dimension 6 \& 8 and dimension 6 derivative corrections. }
\label{F:singlet_R}
\end{figure}

As our analytic results are order-of-magnitude estimates only, and
because we find that SMEFT might provide a good description of the
phase transition in part of parameter space of singlet-extensions, we
also performed numerical scans. The details on the implementation can
be found in \cref{A:scan}.

\Cref{F:singlet} shows the parameter space for a SFO-EWPT in the Higgs-Singlet model (top row) and in SMEFT (bottom row) for $\kappa =4$ (left plots) and $\kappa =2$ (right plots).  The color coding gives the strength of the phase transition $R_N = v_N/T_N$ as a function of scalar vev $s_0$ and mass $m_s$ for the UV theory, and as a function of cutoff scale $\Lambda$ and strength of dimension 8 operator $c_8$ in SMEFT.  In the white area there is no first order phase transition. Comparing the top plots we see that for smaller Higgs-singlet coupling $\kappa$ a SFO-EWPT requires a larger singlet vev. The $3\sigma$ bound on the mixing angle in \cref{mixing} does not constrain the parameter space, but the stronger $2\sigma$ bound -- indicated by the hatched region in the plots -- already cuts almost all parameter space with $\kappa =2$ for a strong phase transition with $R_N >1$. In all of parameter space we find $|A_1|/M \sim 1$ and the constraints \cref{A1_bound} are satisfied.

We have mapped the parameters of the UV theory to those in SMEFT and
vice versa, $(m_s, s_0) \leftrightarrow (\Lambda,c_8)$, using
\cref{mapping,V_SMEFT}.  For example, the solid (dashed) green line
corresponds to $c_8v^2/\Lambda^2 = 0.5 \, (0.3)$ in SMEFT, below which
the EFT expansion (almost) does not converge in the electroweak vacuum
\cref{valid1} and higher order operators are important for a
consistent description. The top plots show the green lines mapped to
the parameters of the Higgs-singlet model.  Further, the red solid
(dashed) lines correspond $R_N =1$ ($R_N =1.4$) in the UV model, and the black solid (dashed) lines to the SMEFT cutoff values of $600$ ($800\,$)GeV. We have included the dimension 6 derivative operator proportional to $c_{\rm kin}$ in the mapping; the magenta lines show the mapping of the equi-$R_N$ for  $c_{\rm kin}=0$. The difference between the red and magenta lines in the bottom plots is only appreciable in the region below the green lines where the EFT description fails. This a posteriori justifies setting $c_{\rm kin}=0$ in  our analytical analysis when focusing on the parameter space of interest.

There are two important lessons to draw from the plots. First, the perturbative UV theory covers only part of parameter space for a SFO-EWPT in SMEFT, in particular the points with $c_8 \ll 1$ have no valid counterpart in the UV theory. Conversely, part of the parameter space for a SFO-EWPT in the UV theory maps to a weak first order phase transition in SMEFT with $R_N <1$.  Second, truncating SMEFT at dimension 6, i.e. with $c_8=0$, does not accurately reproduce the UV results, such as the strength of the 1PT, in SMEFT.

This last point can be better appreciated looking at \cref{F:singlet_R}, which shows the strength of the phase transition in SMEFT $(R_N)_{\rm EFT}$ vs. that in $(R_N)_{\rm UV}$ in the Higgs-singlet model for the points taken from the $\kappa =4$ scan in \cref{F:singlet}.  In the left plot we have included all points with $c_8 v^2/\Lambda^2<0.5$ for which the EFT expansion marginally converges. For perfect agreement between SMEFT and the UV theory all points should lie on the red line in the figure corresponding to $(R_N)_{\rm EFT}=(R_N)_{\rm UV}$. What we see instead is that the blue points for SMEFT with dimension 6 operators reproduce $R_N$-values significantly below that in the full theory. Including dimension 8 operators (orange points), and dimension 8 and dimension 6 derivative operators (green points) improves the matching somewhat.  In the right plot we have only included the points $c_8 v^2/\Lambda^2<0.3$ which have a much better EFT expansion. We see that SMEFT at dimension 6 is still a poor approximation, but the dimension 8 SMEFT gives a much better agreement.  As the effect of dimension 10 operators is small for these points, we expect that the main error here comes from neglecting loop level matching effects and dimension 8 derivative operators.

We end this subsection by noting that our results partially agree with
Ref.~\cite{Damgaard:2015con}, who also compared the predictions of the
Higgs-singlet model and SMEFT truncated at dimension 6. They find that
SMEFT can only give qualitative results for a SFO-EWPT, and only in an
extremely limited region of parameter space; our results indicate a
large overlap region, and also for smaller mixing angles. There are
some differences in how we implemented our numerical analysis. First,
we did not include thermal corrections from diagrams with a singlet
loop, as for the large masses involved, these loops are Boltzmann
suppressed.  Second, we only considered a non-perturbative
Higgs-singlet coupling $\kappa< (4\pi)$, which translates in an upper
bound on the singlet mass for the parameter space of a
SFO-EWPT. Third, we also included the effect of dimension 8
operators. And fourth, and this may be the main cause of the
difference, they restrict all parameters with mass dimension to be
smaller than $M$.  However, in most of the SFO-EWPT parameter space
for which the EFT-expansion is (marginally) valid -- the region above
the green lines in the plots of \cref{F:singlet} -- the ratio
$A^2/M^2 $ is slightly larger than one for (but comfortably smaller
than the bounds \cref{A1_bound} we use).

\subsubsection{Dark sector}
For a dark sector Higgs-singlet type set-up the ratio of Higgs mass to
Higgs vev can be smaller $m_{h_D0}^2/v_D^2 \ll 1$, although for too small
values \cref{dim6_need} the Higgs potential by itself, without the
singlet, can already give a SFO-PT. The requirement for a valid EFT
then becomes
\be
\frac{m_{h_D0}^2}{2v_D^2} \sim \frac38 \frac{\kappa |A_1|^2
   v_D^2}{M^4} \ll  \frac34\frac{|A_1|^2 }{M^2}
 \ee
which allows for a larger separation of scales, that is for smaller
values of $v_D^2/ M^2$ than in the SM, provided $|A_1|^2 / M^2 \ll 1$
is smaller as well.  As discussed in \cref{s:DS} the larger the
separation of scales the smaller the sensitivity to $c_8$.  However,
this is counterbalanced by the increased value of $c_8 \propto
M^2/|A_1|^2$ (see \cref{val_eq2}) in this limit.  We thus expect that
dimension 8 operators will also be important for a dark sector EFT
description of dark Higgs-singlet models.

\subsection{Multifield UV theories}
\label{sec:multifield}

To end this section, we comment on extending the SM with multiple fields. Adding multiple singlets, one can always redefine the fields such that only one singlet direction obtains a vev.  The tree-level matching results will then be the same as in the set-up with a single singlet discussed above.  A UV theory with both singlets and extra Higgs doublets allows to tune parameters in the EFT theory, and maybe reduce the Wilson coefficient of the dimension 8 operator(s). Even if possible, this is in such a limited part of parameter space that it seems  more useful to study the UV model itself than embark on an EFT analysis.

Loop contributions may be enhanced by large-$N$ effects, with $N$ the number of heavy fields, the most interesting case if the dimension 6 operators are enhanced the most (or the perturbutivity constraint weakened). This is not the case in for example $O(N)$-scalar extensions with couplings $\kappa |H|^2 \sum_i s_i^2$; then all one-loop contributions simply pick up a factor $N$ but no further hierarchy between dimension 6 and higher is obtained. It may be be that in generic large $N$ models, with all possible $s_i-s_j$-couplings allowed, the wanted enhancement of the dimension 6 operator is possible.

%%%%%%%%%%%%%%%%%%%%%%%%%%%%%%%%%%%%%%%%%%%%%%%%%%%%%%%%%%%%%%%%%%
%%%%%%%%%%%%%%%%%%%%%%%%%%%%%%%%%%%%%%%%%%%%%%%%%%%%%%%%%%%%%%%%%%
%%%%%%%%%%%%%%%%%%%%%%%%%%%%%%%%%%%%%%%%%%%%%%%%%%%%%%%%%%%%%%%%%%
\section{Conclusion}\label{sec:conclusion}

The nature of the electroweak phase transition is a key question that will be probed by next generation colliders and gravitational wave detectors. It is very attractive then to have a model-independent way of interpreting new results. UV theories that give rise to a strongly first order electroweak phase transition require new degrees of freedom that are relatively light, to obtain the necessary large corrections to the SM Higgs potential. As we have shown in this paper, the lack of a clear separation of scales invalidates the SMEFT description for these set-ups. Unfortunately, updating our knowledge of the electroweak phase diagram thus requires a separate detailed (numerical) study of each (class of) SM extension.

An exception to this are Higgs-singlet models, for which the SMEFT approach can be used to some extent. However, given that SMEFT can only cover part of the interesting parameter space for a first order phase transition, and that accurate agreement with the UV theory is only obtained if dimension 8 operators are included as well, the usefulness of this limited applicability of SMEFT is unclear. As these models have non-zero Higgs mixing angle, they will be further probed by colliders.

Finally, non-renormalizeable dark sectors provide a computationally convenient framework to study gravitational wave production from a strongly first order phase transition. We have derived conditions for when the EFT approach is valid. The applicability to the study of gravitational waves is left for future work.

\section*{Acknowledgments}
MP thanks J. van de Vis for useful discussions. This work
was supported by World Premier International Research Center
Initiative (WPI), MEXT, Japan, and by the Netherlands Organization for Scientific Research (NWO).

%\renewcommand{\theequation}{\thesection.\arabic{equation}}
%\numberwithin{equation}{section}

\appendix

\section{Numerical scan}
\label{A:scan}

In this appendix we detail the input -- the effective potential and parameter values -- for our numerical calculation. To define the effective potential we follow \cite{Delaunay:2007wb,Curtin:2014jma}.  The numerical scans of the phase transition are done with the \texttt{CosmoTransitions} package \cite{Wainwright:2011kj}.

\subsection{Higgs-singlet model}

The one-loop effective potential at finite temperature is $V_{\rm eff} = V_{\rm tree}+V_{\rm CW}+V_T$ with
\begin{align}
  V_{\rm tree} &= \mu_0^2 |H|^2 + \frac12 M^2 s^2 + \lambda_0 |H|^4 +
    \frac14 \lambda_s s^4 + \frac12 \kappa  |H|^2 s^2 - A_1  |H|^2 s + 
                 \frac13 g_s A_1 s^3,\nn \\
  V_{\rm CW} &= \sum_i \frac{n_i}{(8\pi)^2} \[ m_i^4\(\ln
               \(\frac{| m_i^2|}{m_{0i}^2}\)-\frac32\)+2m_i^2m_{0i}^2\] ,\nn\\
  V_T &= \sum_{i={\rm scalar}, A^\parallel} \ n_i T^4 J_{B}
        (\frac{m_{Ti}^2}{T^2})
        +\sum_{i=A^\perp} \ n_i T^4 J_{B} (\frac{m_{i}^2}{T^2})
        +\sum_{i={\rm fermions }} n_j T^4 J_F (\frac{m_{i}^2}{T^2}) .
\end{align}
The thermal functions are given by \cite{Laine:2016hma} \begin{align}
  {J_{B/F}(y^2)} &= \frac{1}{2\pi^2}\int_0^\infty \dd x \, x^2 \ln \(1- s\e^{-\sqrt{x^2 +y^2}}\)
  =
  -s \times  \left\{ \begin{array}{ll}
           \big( \frac{y}{2\pi}\big)^{3/2} \e^{-y}, & y \gg 1 \\
         \frac{  c_0\pi^2}{90}- \frac{c_2 y^2}{24}+ \frac{c_3 y^3}{12\pi} +{\cal O}(y^4), &  y \ll 1
                     \end{array} \right.
 \label{J_lim}                                                                                           
\end{align}
with for bosons $\{s,c_0,c_2,c_3\}=\{1,1,1,1\}$ and for fermions
$\{s,c_0,c_2,c_3\}=\{-1,7/8,1/2,0\}$.

We work in an on-shell renormalization scheme such that
$\partial_i V_{\rm CW}|=\partial_i \partial_j V_{\rm CW}|_{(v,
  s_0)}=0$ -- with $i=h,s$ -- vanishes in the EW vacuum $(h,s)=(v,s_0)$.  The $n_i =\{1,1,3,3,6,-12\}$ gives the degrees of freedom for the Higgs, singlet, goldstones, $Z$, $W$ and top respectively, which together give the dominant one loop contributions. We take the absolute values of $|m_i^2|$ in the log, to assure a real CW potential for negative masses (the Higgs/goldstone masses become negative for small Higgs field values); as argued in \cite{Delaunay:2007wb} the imaginary part of the CW potential is cancelled by an imaginary contribution from the thermal potential, assuring the full potential is real.  The zero temperature masses entering $V_{\rm CW}$ are
\begin{align}
m_W^2 = \frac14 g_2^2 h^2, \quad
m_Z^2 = \frac14 (g_1^2+g_2^2)h^2, \quad
m_t = \frac12 y_t^2 h^2, \quad
m_\chi^2 =\mu_0^2 +\lambda_0 h^2+ \frac12 \kappa s^2 -A_1 s +\eps_\chi.
\end{align}
We included
$\eps_\chi \ll 1$ to keep the log-term well defined in the electroweak vacuum
for the Goldstone bosons (as then $m_{0\chi}^2 = \eps_\chi \neq 0$).
The Higgs and singlet mass eigenstates are obtained by diagonalizing the mass matrix $V_{ij}$
\be
m^2_{h,s} = \frac12\(V_{hh} + V_{ss}
\pm \sqrt{V_{hh}^2 - 2 V_{hh} V_{ss} + V_{ss}^2 + 4 V_{hs}^2}\),
\label{masses}
\ee
where we take the Higgs field to be the lightest mass eigenstate,
corresponding to the minus sign solution above. The
notation is  that $m_{i}^2$ are the higgs field dependent masses, and
$m_{0i}^2 =m_{i}^2|_{(v,s_0)}$ the masses in the EW vacuum.
Explicitly
\be
V_{hh} = -A_1 s + \frac12 \kappa s^2 + 3 \lambda_0 h^2 + \mu_0^2,\quad
V_{ss} = \frac12 h^2 \kappa + 3 \lambda_s s^2 + \mu_s^2 + 2A_1 g_s s,\quad
V_{hs} =h (\kappa s -A_1  )          
\ee
The mass eigenstates are $h_1 = h\cos \theta  +s \sin \theta $
and $h_2 = -h\sin\theta  +s\cos \theta $
with mixing angle
\begin{align}
  \tan \theta &= \frac{y}{1+\sqrt{1+y^2}}, \nn \\
  y &= \frac{2
  V_{hs}}{V_{hh}-V_{ss}}=\frac{2
h (\kappa s - A_1)}{-A_1 s +\frac{1}{2} \kappa s^2 + 3 \lambda _0 h^2 \mu _0 ^2 -\frac{1}{2} h^2 \kappa -3 \lambda _s s^2 - \mu _s^2- 2 A_1 g_s s}
\label{Amixing}
 \end{align}
Since $|\cos \theta|>1/\sqrt{2}$ the $h_1$ mass eigenstate, with mass
$m_h$, is the state with the largest $h$-component \cite{Profumo:2007wc}.

To include leading order infra red thermal corrections from the daisy diagrams we use the thermal masses $m_{iT}^2$ (the leading term in the high $T$ expansion) in $V_T$ for the scalars and longitudinal gauge bosons. At linear order in the high-$T$ expansion this gives the same potential as adding the daisy diagrams separately \cite{Arnold:1992rz}.  As the singlet is heavy, its thermal loop contribution is Boltzmann suppressed and we leave it out from the thermal self-energies. For the transverse gauge d.o.f. and the fermions we can use the zero temperature mass, as these field do not generate daisy corrections.

The thermal self-energies for the scalars are then
\begin{align}
  \Pi_h =\Pi_\chi = T^2\( \frac{g_1^2}{16}  +\frac{3g_2^2}{16}
  +\frac{y_t^2}{4}+\frac{\lambda_0}{4}\), \quad
  \Pi_s = 0 \ . %T^2 \frac{\kappa}{6}
\end{align}
Note that the singlet is heavy and does not contribute to the strong coupling of long wavelength modes, hence its thermal mass is set to zero.
Then $m_{T\chi}^2=m_{\chi}^2 +\Pi_\chi$ and $m_{Ti}^2$ for the singlet/higgs are the eigenvalues of $V_{ab}+ {\rm Diag}(\Pi_h ,\Pi_s )$.
For the longitudinal gauge bosons
\be
m^2_{TW_L} =g_2^2(\frac14 h^2+ \frac{11 }{6} T^2), \quad
m^2_{TZ_L,\gamma_L} = {\rm eigenvalues} \;  {\rm of}\;
\left( \begin{array}{cc}
         g_2^2(\frac14 h^2+\frac{11}{6}T^2) & -\frac{g_1 g_2}{4}h^2 \\
         -\frac{g_1 g_2}{4}h^2 & g_1^2(\frac14 h^2+\frac{11}{6}T^2)
       \end{array}\right)
\ee
where we also include the non-zero thermal photon mass.

We can exchange the parameters $(\lambda_0,A_1,\mu_0^2,M^2)$ for the physical vacuum vevs and masses $(v,s_0,m_{0h}^2,m_{0s}^2)$ using $\partial_s V=\partial_h V|_{(v,s_0)}=0$ and \cref{masses}. The explicit expressions are cumbersome,  but straightforward to implement.
%  Explicitly
% %
% \begin{align}
% \lambda_0 &= \frac{m_{0+}^2 s_0^2 (4 s_0^2 v^2+ Y) -X(Z+4\kappa s_0^2 v^2 )- 
%             16 \lambda_s s_0^6 v^2 }{4 s_0^2 v^2 Y}, \nn \\
%   A_1 &= \frac{m_{0+}^2 X +s_0^2(Z+4 \kappa s_0^2 v^2 -4 \lambda_s X)}{
%   s_0 Y}, \nn \\
%   \mu_0^2 &= - \frac{m_{0+} s_0^2(4s_0^2 v^2 -4 X+Y) -XZ -4 \kappa s_0^2
%             v^2 X -16 s_0^6 v^2 (\kappa+g_s) +2 s_0^4(-2Z+\kappa Y +8
%             \lambda_s X) }{4s_0^2Y}, \nn \\
%   \mu_s^2 &= \frac{m_{0+}^2 \((2s_0^2 + X) 2v^2 -s^2Y\) -XZ - 16
%             \lambda_s s_0^6 v^2 +s_0^2 v^2\(4X(\kappa+2\lambda_s )  -2Z+\kappa Y
%             \) +2s_0^4(4\kappa v^2+\lambda_s Y)}{2 s^2 Y}  
% \end{align}
% %
% with
% %
% \begin{align}
% m_{0\pm}^2 &= m_{0s}^2 \pm m_{0h}^2 ,\nn \\
%   Y &= 4 g_s^2 s_0^4 +4 s_0^2 v^2 +4 g_s s_0^2 v^2 +v^4 ,\nn \\
%   X^2 & =s^4(Y-4s_0^2 v^2),\nn \\
%   Z^2 &= -4 m_{0+}^2 s_0^2v^2 +m_{0-}^2 Y+ 8 m_{0+} v^2(\kappa X +4 s_0^4
%   \lambda_s) -4 s_0^2 v^2 (-4\kappa^2s_0^2 v^2+\kappa^2 Y +8\kappa
%   \lambda_sX +16 \lambda_s^2 s_0^4 )
% \end{align}

%%%%%%%%%%%%%%%%%%%%%%%%%%%%%%%%%%%%%%%%%%%%%%
\subsection{EFT}

The SMEFT potential is given in \cref{V_SMEFT}.
The parameters $a_2,a_4$ are fixed by the vacuum conditions $V_h|_v=0$
and $V_{hh}|_v = m_{0h}^2$ which gives
\begin{align}
  a_2 = -\frac12 m_{0h}^2 + \frac{v^4}{\Lambda^2}+ \frac{2 c_8
  v^6}{\Lambda^4}, \quad a_4v^2=\frac12 m_{0h}^2 - \frac{2v^4}{\Lambda^2}-\frac{3 c_8
  v^6}{\Lambda^4}.
  \end{align}
  The Higgs and Goldstone masses are
  \be
  m_h^2 = a_2+3a_4 h^2 +\frac{5 h^4}{\Lambda^2} + \frac{7c_8
    h^6}{\Lambda^4},\quad  m_\chi^2 = a_2+a_4 h^2 +\frac{ h^4}{\Lambda^2} + \frac{c_8
    h^6}{\Lambda^4}.
  \ee
  We include the corrections from the EFT operators to the
  Higgs/Goldstone self-energies which become 
  \be
  \Pi_{h/\chi}=T^2\( \frac{g_1^2}{16}  +\frac{3g_2^2}{16}
  +\frac{y_t^2}{4} + \frac1{24}(150\frac{c8 h^4}{\Lambda^4}
  +72 \frac{c6 h^2}{\Lambda^2} +12 a_4\).
  \ee
All other SM masses and thermal corrections are the same as in the
singlet model (if we identify $a_2 = \mu_0^2,\, a_4 = \lambda_0$).

\subsection{Parameter values}

\be
g_1=0.377,\quad g_2=0.653,\quad y_t=1, \quad
v=126\,{\rm GeV},\quad m_{h}^2 =125.7\,{\rm GeV}. 
\ee

 \bibliographystyle{jhep} 
\bibliography{refsEFT}

\end{document}